\documentclass[paper]{JHEP3}
\usepackage{epsfig}
\usepackage{amssymb}
\def\beq{\begin{equation}}
\def\eeq{\end{equation}}

\def\beqn{\begin{eqnarray}}
\def\eeqn{\end{eqnarray}}

\def \as{\alpha_{\rm s}}

\def\HW{{\small HERWIG}}
\def\NLO{{\small NLO}}
\def\MC{{\small MC}}

\def\yes{$\checkmark$}
\def\no{$\times$}
\newcommand\sss{\scriptscriptstyle\rm}

\newcommand\MCatNLO{{\rm MC}@{\rm NLO}}
\newcommand\code{\tt}
\newcommand\variable{\tt}
\newcommand\MSbar{{\overline {\rm MS}}}
\newcommand\sinthW{\sin\theta_{\sss W}}
\newcommand\sinsqthW{\sin^2\theta_{\sss W}}
\newcommand\sinfthW{\sin^4\theta_{\sss W}}
\newcommand\pt{p_{\sss T}}
\preprint{
 Cavendish--HEP--08/14\hfill
         }
\title{\boldmath The MC@NLO 3.4 Event Generator%
\footnote{Work supported in part by the UK Science and Technology 
Facilities Council.}}
\author{Stefano Frixione%
  \thanks{On leave of absence from INFN, Sez. di Genova, Italy}\\
  PH Department, TH Unit, CERN, CH-1211 Geneva 23, Switzerland\\
  ITPP, EPFL, CH-1015 Lausanne, Switzerland\\
  E-mail: \email{Stefano.Frixione@cern.ch}}
\author{Bryan R.\ Webber\\
  Cavendish Laboratory, 
  J.J. Thomson Avenue, Cambridge CB3 0HE, U.K.\\
  E-mail: \email{webber@hep.phy.cam.ac.uk}}
\abstract{
This is the user's manual of {$\MCatNLO$} 3.4. This package is a 
practical implementation, based upon the HERWIG event generator,
of the $\MCatNLO$ formalism, which allows one to incorporate NLO QCD 
matrix elements consistently into a parton shower framework.
Processes available in this version include the hadroproduction of
single vector and Higgs bosons, vector boson pairs, heavy
quark pairs, single top, single top in association with a W, 
lepton pairs, and Higgs bosons in association with a $W$ or $Z$. 
Spin correlations are included for all processes except $ZZ$ and $WZ$ 
production. This document is self-contained, but we emphasise the main 
differences with respect to previous versions.
}
\begin{document}

\section{Generalities}
In this document, we briefly describe how to run the $\MCatNLO$ 
package, implemented according to the formalism introduced in 
ref.~\cite{Frixione:2002ik}. 
The production processes now available are listed in tables~\ref{tab:proc}
and~\ref{tab:procdec}. The process codes {\variable IPROC} and the
variables {\variable IV} and {\variable IL}$_\alpha$ will be explained 
below. $H_{1,2}$ represent hadrons (in practice, nucleons or antinucleons).
The information given in refs.~\cite{Frixione:2002ik,Frixione:2003ei}
allows the implementation in MC@NLO of any production process, provided
that the formalism of refs.~\cite{Frixione:1995ms,Frixione:1997np} is
used for the computation of cross sections to NLO accuracy. 
The production matrix elements have been taken from the following references:
vector boson pairs~\cite{Mele:1990bq,Frixione:1992pj,Frixione:1993yp},
heavy quark pairs~\cite{Mangano:1991jk},
Standard Model Higgs~\cite{Dawson:1990zj,Djouadi:1991tk},
single vector boson~\cite{Altarelli:1979ub}, 
lepton pairs~\cite{Aurenche:1980tp},
associated Higgs~\cite{Oleari:2005inprep}
and single-top $s$- and $t$-channel~\cite{Harris:2002md}; 
those for single-top production in association with a $W$ have been 
re-derived and thoroughly compared to those of ref.~\cite{Giele:1995kr}.

This documentation refers to $\MCatNLO$ version 3.4. This version
includes the upgrades of sub-version 3.31, which was not released
officially but was distributed to several experiments.
Single-top production in association with a $W$ has been added since 
sub-version 3.31, including spin correlations. Top hadron
decays (at the leading order) with spin correlations are now included.
New Monte Carlo subtraction terms have been implemented in single-top 
production (all channels) -- they were already implemented in $Q\bar{Q}$
production in sub-version 3.31, and they coincide with the old ones
for all other processes. The automatic assignment of $\Lambda_{\sss QCD}$
in conjunction with LHAPDF has been improved. As a standalone package, 
$\MCatNLO$ version 3.4 should be easier to link to any external libraries
(such as Root, for which we provide a Fortran interface~\cite{WVroot})
and to recent versions of LHAPDF. For precise details of version changes, 
see app.~\ref{app:newver}-\ref{app:newverg}.

\subsection{Citation policy\label{sec:cites}}
When using $\MCatNLO$, please cite ref.~\cite{Frixione:2002ik}. In addition 
to ref.~\cite{Frixione:2002ik}, if $t\bar{t}$ or $b\bar{b}$ events are 
generated, please also cite ref.~\cite{Frixione:2003ei}; if $s$- or 
$t$-channel single-top events are generated, please also cite 
ref.~\cite{Frixione:2005vw}; if $Wt$ single-top events are generated,
please also cite ref.~\cite{Frixione:2008yi}. The current user manual,
or any other user manuals relevant to past versions, should not be cited
unless the relevant papers mentioned above are cited too.

\subsection{Mode of operation\label{sec:oper}}
In the case of standard MC, a hard kinematic configuration is
generated on a event-by-event basis, and it is subsequently showered 
and hadronized. In the case of $\MCatNLO$, all of the hard kinematic
configurations are generated in advance, and stored in a file 
(which we call {\em event file} -- see sect.~\ref{sec:evfile}); 
the event file is then read by \HW, which showers and hadronizes each 
hard configuration. Since version 2.0, the events are 
handled by the ``Les Houches'' generic user process 
interface~\cite{Boos:2001cv} (see ref.~\cite{Frixione:2003ei} for 
more details). Therefore, in $\MCatNLO$ the reading of a 
hard configuration from the event file is equivalent to the generation 
of such a configuration in a standard MC.

The signal to \HW\ that configurations should be read from an event file using
the Les Houches interface is a negative value of the process code {\variable
IPROC}; this accounts for the negative values in tables~\ref{tab:proc}
and~\ref{tab:procdec}. 
In the case of heavy quark pair, Higgs, Higgs in association with a $W$ 
or $Z$, and lepton pair (through $Z/\gamma^*$ exchange) production, the codes 
are simply the negative of those for the corresponding standard \HW\ MC
processes. Where possible, this convention will be adopted for additional
$\MCatNLO$ processes.  Consistently with what happens in standard \HW, by
subtracting 10000 from {\variable IPROC} one generates the same processes as
in tables~\ref{tab:proc} and~\ref{tab:procdec}, but eliminates the underlying 
event\footnote{The same effect can be achieved by setting the 
{\scriptsize HERWIG} parameter {\variable PRSOF} $=0$.}.
 
Higgs decays are controlled in the same way as in \HW, that is by adding
{\variable -ID} to the process code. The conventions for {\variable ID}
are the same as in \HW, namely {\variable ID} $=1\ldots 6$ for 
$u\bar{u}\ldots t\bar{t}$; $7$, $8$, $9$ for $e^+e^-$, $\mu^+\mu^-$, 
$\tau^+\tau^-$; $10, 11$ for $W^+W^-, ZZ$; and 12 for $\gamma\gamma$. 
Furthermore, {\variable ID} $=0$ gives quarks of all flavours, and 
{\variable ID} $=99$ gives all decays. It should be stressed that 
the event file does not contain the Higgs decay products, and therefore
is independent of the value of {\variable ID}; the decay is
dealt with by \HW.\footnote{In the current version of \HW\ (6.510),
  spin correlations between the products of Higgs decays are
  neglected. In version 6.520, to be released shortly, spin
  correlations in decays to vector boson pairs are included.  
  Please check the Fortran \HW\ wiki at
  http://projects.hepforge.org/fherwig/trac/report
  for pre-release reports on this and other improvements.}

Process codes {\variable IPROC}=$-1360-${\variable IL} and 
$-1370-${\variable IL} do not have an analogue in \HW; they are the 
same as $-1350-${\variable IL}, except for the fact that only a $Z$ or a
$\gamma^*$ respectively is exchanged. The value of {\variable IL} determines
the lepton identities, and the same convention as in \HW\ is adopted:
{\variable IL}=$1,\ldots,6$ for $l_{\rm IL}=e,\nu_e,\mu,\nu_\mu,\tau,\nu_\tau$
respectively. At variance with \HW, {\variable IL} cannot be set equal to
zero. Process codes {\variable IPROC}=$-1460-${\variable IL} and
$-1470-${\variable IL} are the analogue of \HW\ $1450+${\variable IL};
in \HW\, either $W^+$ or $W^-$ can be produced, whereas MC@NLO treats
the two vector bosons separately. For these processes, as in \HW,
{\variable IL}=$1,2,3$ for $l_{\rm IL}=e,\mu,\tau$, but again
the choice ${\variable IL}=0$ is not allowed.

\begin{table}[h!]
\centering
\begin{tabular}{|c|c|c|c|c|l|}\hline
{\variable IPROC} & {\variable IV} & {\variable IL}$_1$ & {\variable IL}$_2$ & 
 Spin & Process \\\hline
 --1350--{\variable IL} & & & &\yes &
 $H_1 H_2\to (Z/\gamma^*\to) l_{\rm IL}\bar{l}_{\rm IL}+X$\\\hline
 --1360--{\variable IL} & & & &\yes &
 $H_1 H_2\to (Z\to) l_{\rm IL}\bar{l}_{\rm IL}+X$\\\hline
 --1370--{\variable IL} & & & &\yes &
 $H_1 H_2\to (\gamma^*\to) l_{\rm IL}\bar{l}_{\rm IL}+X$\\\hline
 --1460--{\variable IL} & & & &\yes &
 $H_1 H_2\to (W^+\to) l_{\rm IL}^+\nu_{\rm IL}+X$\\\hline
 --1470--{\variable IL} & & & &\yes &
 $H_1 H_2\to (W^-\to) l_{\rm IL}^-\bar{\nu}_{\rm IL}+X$\\\hline
 --1396 & & & &\no &
 $H_1 H_2\to \gamma^*(\to \sum_i f_i\bar{f}_i)+X$\\\hline
 --1397 & & & &\no &
 $H_1 H_2\to Z^0+X$\\\hline
 --1497 & & & &\no &
 $H_1 H_2\to W^+ +X$\\\hline
 --1498 & & & &\no &
 $H_1 H_2\to W^- +X$\\\hline
 --1600--{\variable ID} & & & & &
 $H_1 H_2\to H^0+X$\\\hline
 --1705 & & & & &
 $H_1 H_2\to b\bar{b}+X$\\\hline
 --1706 & & 7 & 7 & \no &
 $H_1 H_2\to t\bar{t}+X$\\\hline
 --2000--{\variable IC} & & 7 & & \no &
 $H_1 H_2\to t/\bar{t}+X$\\\hline
 --2001--{\variable IC} & & 7 & & \no &
 $H_1 H_2\to \bar{t}+X$\\\hline
 --2004--{\variable IC} & & 7 & & \no &
 $H_1 H_2\to t+X$\\\hline
 --2030 & & 7 & 7 & \no &
 $H_1 H_2\to tW^-/\bar{t}W^+ +X$\\\hline
 --2031 & & 7 & 7 & \no &
 $H_1 H_2\to \bar{t}W^+ +X$\\\hline
 --2034 & & 7 & 7 & \no &
 $H_1 H_2\to tW^- +X$\\\hline
 --2600--{\variable ID} & 1 & 7 & &\no &
 $H_1 H_2\to H^0 W^+ +X$\\\hline
 --2600--{\variable ID} & 1 & $i$ & &\yes &
 $H_1 H_2\to H^0 (W^+\to)l_i^+\nu_i +X$\\\hline
 --2600--{\variable ID} & -1 & 7 & &\no &
 $H_1 H_2\to H^0 W^- +X$\\\hline
 --2600--{\variable ID} & -1 & $i$ & &\yes &
 $H_1 H_2\to H^0 (W^-\to)l_i^-\bar{\nu}_i +X$\\\hline
 --2700--{\variable ID} & 0 & 7 & &\no &
 $H_1 H_2\to H^0 Z +X$\\\hline
 --2700--{\variable ID} & 0 & $i$ & &\yes &
 $H_1 H_2\to H^0 (Z\to)l_i\bar{l}_i +X$\\\hline
 --2850 & & 7 & 7 & \no &
 $H_1 H_2\to W^+W^-+X$\\\hline
 --2860 & & 7 & 7 & \no &
 $H_1 H_2\to Z^0Z^0+X$\\\hline
 --2870 & & 7 & 7 & \no &
 $H_1 H_2\to W^+Z^0+X$\\\hline
 --2880 & & 7 & 7 & \no &
 $H_1 H_2\to W^-Z^0+X$\\\hline
\end{tabular}
\caption{\label{tab:proc} 
Some of the processes implemented in $\MCatNLO~3.4$ (see also
table~\ref{tab:procdec}). $H_{1,2}$ represent nucleons or
antinucleons. $H^0$ denotes the Standard Model Higgs boson 
and the value of {\variable ID} controls its decay, as described in 
the \HW\ manual and in the text. The values of {\variable IV}, {\variable IL},
{\variable IL}$_1$, and {\variable IL}$_2$ control the identities of vector
bosons and leptons, as described in the text. In single-$t$ production, the
value of {\variable IC} controls the production processes ($s$- and/or
$t$-channel), as described in the text. 
For more details on $Wt$ production, see sect.~\ref{sec:Wt}.
{\variable IPROC}--10000 generates the same processes as 
{\variable IPROC}, but eliminates the underlying event. A void entry 
indicates that the corresponding variable is unused. The `Spin' column 
indicates whether spin correlations in vector boson or top decays are 
included (\yes), neglected (\no) or absent (void entry); when included,
spin correlations are obtained by direct integration of the relevant
NLO matrix elements. Spin correlations in Higgs decays to vector boson
pairs (e.g.\ $H^0\to W^+W^-\to l^+\nu l^-\bar{\nu}$) are included in
\HW\ versions 6.520 and higher.  }
\end{table}
\begin{table}[h!]
\centering
\begin{tabular}{|c|c|c|c|c|l|}\hline
{\variable IPROC} & {\variable IV} & {\variable IL}$_1$ & {\variable IL}$_2$ & 
 Spin & Process \\\hline
 --1706 & & $i$ & $j$ & \yes &
 $H_1 H_2\to (t\to)b_k f_if^\prime_i 
             (\bar{t}\to)\bar{b}_l f_jf^\prime_j+X$\\\hline
 --2000--{\variable IC} & & $i$ & & \yes &
 $H_1 H_2\to (t\to)b_k f_if^\prime_i/
             (\bar{t}\to)\bar{b}_k f_if^\prime_i+X$\\\hline
 --2001--{\variable IC} & & $i$ & & \yes &
 $H_1 H_2\to (\bar{t}\to)\bar{b}_k f_if^\prime_i+X$\\\hline
 --2004--{\variable IC} & & $i$ & & \yes &
 $H_1 H_2\to (t\to)b_k f_if^\prime_i+X$\\\hline
 --2030 & & $i$ & $j$ & \yes &
 $H_1 H_2\to (t\to)b_k f_if^\prime_i (W^-\to)f_jf^\prime_j/$\\
        & & & & & 
 $\phantom{H_1 H_2\to\,}
 (\bar{t}\to)\bar{b}_k f_if^\prime_i(W^+\to)f_jf^\prime_j+X$\\\hline
 --2031 & & $i$ & $j$ & \yes &
 $H_1 H_2\to 
 (\bar{t}\to)\bar{b}_k f_if^\prime_i(W^+\to)f_jf^\prime_j+X$\\\hline
 --2034 & & $i$ & $j$ & \yes &
 $H_1 H_2\to (t\to)b_k f_if^\prime_i (W^-\to)f_jf^\prime_j+X$\\\hline
 --2850 & & $i$ & $j$ & \yes &
 $H_1 H_2\to (W^+\to)l_i^+\nu_i (W^-\to)l_j^-\bar{\nu}_j +X$\\\hline
\end{tabular}
\caption{\label{tab:procdec} 
Some of the processes implemented in $\MCatNLO~3.4$ (see also
table~\ref{tab:proc}). $H_{1,2}$ represent nucleons or
antinucleons. For more details on $Wt$ production, see 
sect.~\ref{sec:Wt}. Spin correlations for the processes in this
table are implemented according to the method presented 
in ref.~\cite{Frixione:2007zp}. 
$b_\alpha$ ($\bar{b}_\alpha$) can either denote a $b$ (anti)quark
or a generic down-type (anti)quark. $f_\alpha$ and $f^\prime_\alpha$
can denote a (anti)lepton or an (anti)quark. See sects.~\ref{sec:xsecs} 
and~\ref{sec:decay} for fuller details. }
\end{table}

The lepton pair processes {\variable IPROC}=$-1350-${\variable IL},
$\ldots$, $-1470-${\variable IL} include spin correlations when
generating the angular distributions of the
produced leptons. However, if spin correlations are not an
issue, the single vector boson production processes
{\variable IPROC}= $-$1396,$-$1397,$-$1497,$-$1498 can be used,
in which case the vector boson decay products are distributed
(by \HW, which then generates the decays) according to phase space.

There are a number of other differences between the lepton pair and single
vector boson processes. The latter do not feature the $\gamma$--$Z$
interference terms. Also, their cross sections are fully inclusive in the
final-state fermions resulting from $\gamma^*$, $Z$ or $W^\pm$.  The user can
still select a definite decay mode using the variable {\variable MODBOS} (see
sect.~\ref{sec:decay}), but the relevant branching ratio will {\em not} be
included by MC@NLO.

In NLO computations for single-top production, it is customary to
distinguish between three production mechanisms, conventionally
denoted as $s$ channel, $t$ channel, and $Wt$ mode. Starting from the
current version 3.4, all three mechanisms are implemented in $\MCatNLO$;
$s$- and $t$-channel single top production correspond to setting 
{\variable IC}=$10$ and {\variable IC}=$20$ respectively. 
For example, according to tables~\ref{tab:proc} and~\ref{tab:procdec}, 
$t$-channel single-$\bar{t}$ events will be generated by entering 
{\variable IPROC}=$-2021$. These two channels can also be simulated 
simultaneously (by setting {\variable IC}=$0$). We point out that
$Wt$ cross section is ill-defined beyond the leading order in QCD.
See sect.~\ref{sec:Wt} for more details.

In the case of vector boson pair production, the process codes are the 
negative of those adopted in $\MCatNLO$ 1.0 (for which the Les Houches 
interface was not yet available), rather than those of standard \HW.

Furthermore, in the case of $t\bar{t}$, single-$t$, $H^0W^\pm$, 
$H^0Z$ and $W^+W^-$ production, the value of {\variable IPROC} alone
may not be sufficient to fully determine the process type (including
decay products), and variables
{\variable IV}, {\variable IL}$_1$, and {\variable IL}$_2$ are also
needed (see tables~\ref{tab:proc} and~\ref{tab:procdec}).  
In the case of top decays (and of the decay of the hard $W$ 
in $Wt$ production), the variables {\variable IL}$_1$ 
and {\variable IL}$_2$ have a more extended range of values 
than that of the variable {\variable IL}, which is relevant to
lepton pair production and to which they are analogous
(notice, however, that in the latter case {\variable IL} is not 
an independent variable, and its value is included via {\variable
IPROC}). In addition, {\variable IL}$_\alpha$=7 implies that spin
correlations for the decay products of the corresponding particle are not
taken into account, as indicated in table~\ref{tab:proc}.
More details are given in sect.~\ref{sec:decay}.

Apart from the above differences, $\MCatNLO$ and \HW\ {\em behave in exactly
the same way}. Thus, the available user's analysis routines can be 
used in the case of $\MCatNLO$. One should recall, however, that
$\MCatNLO$ always generates some events with negative weights (see
refs.~\cite{Frixione:2002ik}); therefore, the correct distributions 
are obtained by summing weights with their signs (i.e., the absolute 
values of the weights must {\em NOT} be used when filling the histograms).

With such a structure, it is natural to create two separate executables,
which we improperly denote as \NLO\ and \MC. The former has the sole scope
of creating the event file; the latter is just \HW, augmented by
the capability of reading the event file.

\subsection{Package files\label{sec:packfile}}
The package consists of the following files:

\begin{itemize}
\item {\bf Shell utilities}\\
    {\code MCatNLO.Script}\\
    {\code MCatNLO.inputs}\\
    {\code MCatNLO\_dyn.Script}\\
    {\code MCatNLO\_rb.inputs}\\
    {\code Makefile}\\
    {\code Makefile\_dyn}

\item {\bf Utility codes}\\
    {\code MEcoupl.inc}\\ 
    {\code alpha.f}\\ 
    {\code dummies.f}\\ 
    {\code linux.f}\\ 
    {\code mcatnlo\_date.f}\\  
    {\code mcatnlo\_hbook.f}\\  
    {\code mcatnlo\_helas2.f}\\  
    {\code mcatnlo\_hwdummy.f}\\ 
    {\code mcatnlo\_int.f}\\
    {\code mcatnlo\_libofpdf.f}\\ 
    {\code mcatnlo\_mlmtolha.f}\\  
    {\code mcatnlo\_mlmtopdf.f}\\ 
    {\code mcatnlo\_pdftomlm.f}\\ 
    {\code mcatnlo\_str.f}\\ 
    {\code mcatnlo\_uti.f}\\ 
    {\code mcatnlo\_utilhav4.f}\\ 
    {\code mcatnlo\_uxdate.c}\\
    {\code rbook\_be.cc}\\
    {\code rbook\_fe.f}\\
    {\code sun.f}\\ 
    {\code trapfpe.c}

\item {\bf General \HW\ routines}\\
    {\code mcatnlo\_hwdriver.f}\\ 
    {\code mcatnlo\_hwlhin.f}

\item {\bf Process-specific codes}\\
    {\code mcatnlo\_hwan{\em xxx}.f}\\
    {\code mcatnlo\_hwan{\em xxx}\_rb.f}\\
    {\code mcatnlo\_hgmain.f}\\
    {\code mcatnlo\_hgxsec.f}\\
    {\code mcatnlo\_llmain.f}\\
    {\code mcatnlo\_llxsec.f}\\
    {\code mcatnlo\_qqmain.f}\\
    {\code mcatnlo\_qqxsec.f}\\
    {\code mcatnlo\_sbmain.f}\\
    {\code mcatnlo\_sbxsec.f}\\
    {\code mcatnlo\_stmain.f}\\
    {\code mcatnlo\_stxsec.f}\\
    {\code mcatnlo\_vbmain.f}\\
    {\code mcatnlo\_vbxsec.f}\\
    {\code mcatnlo\_vhmain.f}\\
    {\code mcatnlo\_vhxsec.f}\\
    {\code mcatnlo\_wtmain\_dr.f}\\
    {\code mcatnlo\_wtmain\_ds.f}\\
    {\code mcatnlo\_wtxsec\_dr.f}\\
    {\code mcatnlo\_wtxsec\_ds.f}\\
    {\code hgscblks.h}\\
    {\code hvqcblks.h}\\
    {\code llpcblks.h}\\
    {\code stpcblks.h}\\
    {\code svbcblks.h}\\
    {\code vhgcblks.h}
\end{itemize}
These files can be downloaded from the web page:\\
$\phantom{aaaaaaaa}$%
{\code http://www.hep.phy.cam.ac.uk/theory/webber/MCatNLO}\\
The files {\code mcatnlo\_hwan{\em xxx}.f} (which use a version
of HBOOK written by M.~Mangano that outputs plots in TopDrawer format) 
and {\code mcatnlo\_hwan{\em xxx}\_rb.f} (which use front-end Fortran
routines written by W.~Verkerke~\cite{WVroot} for filling histograms in 
Root format) are sample \HW\ analysis routines. They are provided here 
to give the user a ready-to-run package, but they should be replaced 
with appropriate codes according to the user's needs. Examples of
how to use these analysis files in $\MCatNLO$ are given in the
(otherwise identical) {\code MCatNLO.inputs} and 
{\code MCatNLO\_rb.inputs} files (see sect.~\ref{sec:running}
for more details on input cards).

In addition to the files listed above, the user will need a
version of the \HW\ code
\cite{Marchesini:1992ch,Corcella:2001bw,Corcella:2002jc}.
As stressed in 
ref.~\cite{Frixione:2002ik}, for the $\MCatNLO$ we do not
modify the existing (LL) shower algorithm. However, since $\MCatNLO$
versions 2.0 and higher make use of the Les Houches interface,
first implemented in \HW\ 6.5, the version must be 6.500 or higher.
On most systems, users will need to delete the dummy  subroutines 
{\small UPEVNT}, {\small UPINIT}, {\small PDFSET} and {\small STRUCTM}
from the standard  \HW\ package, to permit linkage of the corresponding
routines from the $\MCatNLO$ package. As a general rule, the user is
strongly advised to use the most recent version of \HW\ (currently
6.510 -- with versions lower than 6.504 problems can be found in attempting 
to specify the decay modes of single vector bosons through the variable
{\variable MODBOS}. Also, crashes in the shower phase have been reported 
when using \HW\ 6.505, and we therefore recommend not to use that
version).

\subsection{Working environment}
We have written a number of shell scripts and a {\code Makefile} (all
listed under {\bf Shell utilities} above) which will simplify the use of
the package considerably. In order to use them, the computing system
must support {\code bash} shell, and {\code gmake}\footnote{For Macs 
running under OSX v10 or higher, {\code make} can be used instead of 
{\code gmake}.}. 
Should they be unavailable on the user's computing system, the compilation 
and running of $\MCatNLO$ requires more detailed instructions; in this case,
we refer the reader to app.~\ref{app:instr}. This appendix will serve also as
a reference for a more advanced use of the package.

\subsection{Source and running directories}
We assume that all the files of the package sit in the same directory,
which we call the {\em source directory}. When creating the executable, 
our shell scripts determine the type of operating system, and create a
subdirectory of the source directory, which we call the {\em running 
directory}, whose name is {\variable Alpha}, {\variable Sun}, {\variable
Linux}, or {\variable Darwin}, depending on the operating system.  
If the operating system is not known by our scripts, the name of the 
working directory is {\variable Run}. The running directory contains all 
the object files and executable files, and in general all the files produced
by the $\MCatNLO$ while running.  It must also contain the relevant grid files
(see sect.~\ref{sec:pdfs}), or links to them, if the library of parton
densities provided with the $\MCatNLO$ package is used.

\section{Prior to running\label{sec:priors}}
Before running the code, the user must be aware of the fact that the files:\\
$\phantom{aaa}${\code mcatnlo\_hwdriver.f}\\ 
$\phantom{aaa}${\code mcatnlo\_hwlhin.f}\\
$\phantom{aaa}${\code mcatnlo\_hwan{\em xxx}.f}\\
$\phantom{aaa}${\code mcatnlo\_hwan{\em xxx}\_rb.f}\\
contain the statement {\code INCLUDE HERWIG65.INC}, which indicates
that the code will link to \HW\ version 6.500 or higher, for the
reasons explained above. In the current MC@NLO release, the file
{\code HERWIG65.INC} contains the statement\\
$\phantom{aaaaaa}${\code INCLUDE 'herwig6510.inc'}\\
We do not assume that the user will adopt version 6.510, which is
the latest release of \HW; for this reason, the user will in general have 
to edit the file {\code HERWIG65.INC}, and change the statement
above into\\
$\phantom{aaaaaa}${\code INCLUDE 'herwig65nn.inc'}\\
with {\code 65nn} the \HW\ version chosen by the user (this must be
consistent with the value of the input parameter {\variable HERWIGVER},
see sects.~\ref{sec:running} and~\ref{sec:scrvar}).

The file {\code mcatnlo\_hwdriver.f} contains a set of read statements,
which are necessary for the \MC\ to get the input parameters (see
sect.~\ref{sec:running} for the input procedure); these read
statements must not be modified or eliminated. Also, {\code
mcatnlo\_hwdriver.f} calls the \HW\ routines which
perform showering, hadronization, decays (see sect.~\ref{sec:decay} 
for more details on this issue), and so forth; the user can
freely modify this part, as customary in \MC\ runs. Finally, the sample
codes {\code mcatnlo\_hwan{\em xxx}.f} and
{\code mcatnlo\_hwan{\em xxx}\_rb.f} contain analysis-related routines:
these files must be replaced by files which contain the user's analysis 
routines. We point out that, since version 2.0, the {\code Makefile} need not
be edited any longer, since the corresponding operations are now 
performed by setting script variables (see sect.~\ref{sec:scrvar}).

\section{Running\label{sec:running}}
It is straightforward to run $\MCatNLO$. First, edit\footnote{See
below for comments on {\code MCatNLO\_rb.inputs}}\\
$\phantom{aaa}${\code MCatNLO.inputs}\\
and write there all the input parameters (for the complete list 
of the input parameters, see sect.~\ref{sec:scrvar}). As the last
line of the file {\code MCatNLO.inputs}, write\\
$\phantom{aaa}${\code runMCatNLO}\\
Finally, execute {\code MCatNLO.inputs} from the {\code bash} shell.
This procedure will create the \NLO\ and \MC\ executables, and run them
using the inputs given in {\code MCatNLO.inputs}, which guarantees
that the parameters used in the \NLO\ and \MC\ runs are consistent.
Should the user only need to create the executables without running
them, or to run the \NLO\ or the \MC\ only, he/she should replace the
call to {\code runMCatNLO} in the last line of {\code MCatNLO.inputs}
by calls to\\
$\phantom{aaa}${\code compileNLO}\\
$\phantom{aaa}${\code compileMC}\\
$\phantom{aaa}${\code runNLO}\\
$\phantom{aaa}${\code runMC}\\
which have obvious meanings. We point out that the command {\code runMC}
may be used with {\variable IPROC}=1350+{\variable IL}, 1450+{\variable IL}, 
1600+{\variable ID}, 1699, 1705, 1706, 2000--2008, 2600+{\variable ID}, 2699, 
2700+{\variable ID}, 2799, 2800, 2810, 2815, 2820, 2825 to generate 
$Z/\gamma^*$, $W^\pm$, Higgs, $b\bar{b}$, $t\bar{t}$, single top, $H^0W$,
$H^0Z$, and vector boson pair events with standard \HW\ (see the \HW\ 
manual for more details).

We stress that the input parameters are not solely related to
physics (masses, CM energy, and so on); there are several of them
which control other things, such as the number of events generated.
These must also be set by the user, according to his/her needs:
see sect.~\ref{sec:scrvar}.

Two such variables are {\variable HERWIGVER} and {\variable HWUTI},
which were moved in version 2.0 from the {\code Makefile} to
{\code MCatNLO.inputs}. The former variable must be set equal 
to the object file name of the version of \HW\ currently adopted 
(matching the one whose common blocks are included in the files
mentioned in sect.~\ref{sec:priors}). The variable {\variable HWUTI} 
must be set equal to the list of object files that the user needs in 
the analysis routines.

The sample input file {\code MCatNLO.inputs} provided in this package 
is relevant to $t\bar{t}$ production and subsequent $t$ and $\bar{t}$  
leptonic decays. 
Similar sample inputs are given in the file {\code MCatNLO\_rb.inputs}, 
which is identical to the former, except that at the end of the MC run
an output file in Root format will be produced (as opposed to the
output file in TopDrawer format produced by {\code MCatNLO.inputs});
for this to happen, the user will have to edit {\code MCatNLO\_rb.inputs}
in order to insert the path to the Root libraries for the machine 
on which the run is performed (shell variables {\code EXTRAPATHS} and
{\code INCLUDEPATHS}). We stress that, apart from the differences in
the output formats, {\code MCatNLO.inputs} and {\code MCatNLO\_rb.inputs}
have exactly the same meaning. Thus, although for the sake of brevity
we shall often refer only to {\code MCatNLO.inputs} in this manual,
all the issues concerning the inputs apply to {\code MCatNLO\_rb.inputs}
as well.

If the shell scripts are not used to run the codes, the inputs are
given to the \NLO\ or \MC\ codes during an interactive talk-to phase;
the complete sets of inputs for our codes are reported in 
app.~\ref{app:input} for vector boson pair production.

\subsection{Parton densities\label{sec:pdfs}}
Since the knowledge of the parton densities (PDFs) is necessary in
order to get the physical cross section, a PDF library must be
linked. The possibility exists to link the (now obsolete) CERNLIB 
PDF library (PDFLIB), or its replacement LHAPDF~\cite{Whalley:2005nh};
however, we also provide a self-contained PDF library with this package, 
which is faster than PDFLIB, and contains PDF sets released after the 
last and final PDFLIB version (8.04; most of these sets are now included 
in LHAPDF). A complete list of the PDFs available in our PDF library can 
be downloaded from the MC@NLO web page. The user may link one of the three
PDF libraries; all that is necessary is to set the variable {\variable
PDFLIBRARY} (in the file {\code MCatNLO.inputs}) equal to {\variable
THISLIB} if one wants to link to our PDF library, and equal to
{\variable PDFLIB} or to {\variable LHAPDF} if one wants to link 
to PDFLIB or to LHAPDF.  Our PDF library collects 
the original codes, written by the authors of the PDF fits;
as such, for most of the densities it needs to read the files which
contain the grids that initialize the PDFs. These files, which can
also be downloaded from the $\MCatNLO$ web page, must either be copied 
into the running directory, or defined in the running directory as logical
links to the physical files (by using {\code ln -sn}). We stress that if
the user runs $\MCatNLO$ with the shell scripts, the logical links will
be created automatically at run time.

As stressed before, consistent inputs must be given to the \NLO\ and
\MC\ codes. However, in ref.~\cite{Frixione:2002ik} we found that the
dependence upon the PDFs used by the MC is rather weak. So one may
want to run the \NLO\ and \MC\ adopting a regular NLL-evolved set in the
former case, and the default \HW\ set in the latter (the advantage is
that this option reduces the amount of running time of the \MC). In
order to do so, the user must set the variable {\variable HERPDF}
equal to {\variable DEFAULT} in the file {\code MCatNLO.inputs};
setting {\variable HERPDF=EXTPDF} will force the \MC\ to use the same
PDF set as the \NLO\ code.

Regardless of the PDFs used in the \MC\ run, users must delete the dummy 
PDFLIB routines {\small PDFSET} and {\small STRUCTM} from \HW, as
explained earlier.

\subsubsection{LHAPDF\label{sec:lhapdf}}
As mentioned above, by setting {\variable THISLIB}$=${\variable LHAPDF}
in the input file the code is linked to the LHAPDF library. By default,
$\MCatNLO$ will link to the static LHAPDF library. If one wants to link
to the dynamic LHAPDF library (which will produce a smaller executable
but otherwise identical results), one needs to replace\\
$\phantom{aa}${\variable . \$thisdir/MCatNLO.Script}\\
in {\code MCatNLO.inputs} with\\
$\phantom{aa}${\variable . \$thisdir/MCatNLO\_dyn.Script}\\
In order for the {\code Makefile} (or {\code Makefile\_dyn}, in the
case of dynamic libraries) to be able to find the LHAPDF library,
the variable {\variable LHAPATH} in {\code MCatNLO.inputs} should
be set equal to the name of the directory where the local 
version of LHAPDF is installed. This is typically the name of
the directory where one finds the files {\code libLHAPDF.a}
and {\code libLHAPDF.so}, except for the final {\code /lib} in
the directory name.

As is well known, a given PDF set has a preferred value of 
$\Lambda_{\sss QCD}$,
which should be used in the computation of short-distance cross sections.
Upon setting {\variable LAMBDAFIVE} in {\code MCatNLO.inputs} equal
to a negative value, this choice is made automatically. However, when
linking to PDFLIB or LHAPDF, the code has to rely on the value
$\Lambda_{\sss QCD}$ stored (by the PDF libraries) in a common block.
This is far from ideal, since $\Lambda_{\sss QCD}$ is not a physical
parameter, and in particular is dependent upon the form adopted for
$\as$, which may not be the same as that used in MC@NLO.
Starting from version 3.4, the above automatic choice has been
rendered more solid in the case of a linkage to LHAPDF; the code
now reads the value of $\as(M_Z)$ (i.e., of a physical quantity) 
from the PDF library, and converts it into a value for $\Lambda_{\sss QCD}$
using the form of $\as(Q^2)$ used internally in MC@NLO. MC@NLO will
print out on the standard output when running the NLO code
({\variable FPREFIXNLO.log} if using the scripts) the value of
$\Lambda_{\sss QCD}$ used in the computation. Such a value is now 
expected to be quite close to that listed under the column labeled
with $\Lambda_{\sss QCD}^{(5)}$(MeV) on our PDF library manual
(which can be found on the MC@NLO web page).

Version 3.4 of MC@NLO has been tested to link and run with several
versions of LHAPDF. In particular, the user is not supposed to edit 
the {\code Makefile} if linking with LHAPDF version 5.0 or higher.
If one is interested into linking with earlier versions of LHAPDF,
then one must replace the string {\code mcatnlo\_uti.o} in the variable
{\variable LUTIFILES} in the {\code Makefile} (or {\code Makefile\_dyn}, 
in the case of dynamic libraries) with the string\\ 
{\code mcatnlo\_utilhav4.o}.

\subsection{Event file\label{sec:evfile}}
The \NLO\ code creates the event file. In order to do so, it goes through
two steps; first it integrates the cross sections (integration step),
and then, using the information gathered in the integration step, 
produces a set of hard events (event generation step). Integration and
event generation are performed with a modified version of the 
{\small SPRING-BASES} package~\cite{Kawabata:1995th}.

We stress that the events stored in the event file just contain the
partons involved in the hard suprocesses. Owing to the modified subtraction
introduced in the $\MCatNLO$ formalism (see ref.~\cite{Frixione:2002ik}) 
they do not correspond to pure NLO configurations, and should not be 
used to plot physical observables. Parton-level observables must be
reconstructed using the fully-showered events.

The event generation step necessarily follows the integration step;
however, for each integration step one can have an arbitrary number of
event generation steps, i.e., an arbitrary number of event files.
This is useful in the case in which the statistics accumulated 
with a given event file is not sufficient.

Suppose the user wants to create an event file; editing {\code
MCatNLO.inputs}, the user sets {\variable BASES=ON}, to enable the
integration step, sets the parameter {\variable NEVENTS} equal to
the number of events wanted on tape, and runs the code; the
information on the integration step (unreadable to the user, but
needed by the code in the event generation step) is written on files
whose name begin with {\variable FPREFIX}, a string the user sets
in {\code MCatNLO.inputs}; these files (which we denotes as {\em data
files}) have extensions {\code .data}. The name of the event file is 
{\variable EVPREFIX.events}, where {\variable EVPREFIX} is again a 
string set by the user.

Now suppose the user wants to create another event file, to increase
the statistics. The user simply sets {\variable BASES=OFF}, since 
the integration step is not necessary any longer (however, the data
files must not be removed: the information
stored there is still used by the \NLO\ code); changes the string
{\variable EVPREFIX} (failure to do so overwrites the existing event
file), while keeping {\variable FPREFIX} at the same value as before;
and changes the value of {\variable RNDEVSEED} (the random number
seed used in the event generation step; failure to do so results in
an event file identical to the previous one); the number {\variable
NEVENTS} generated may or may not be equal to the one chosen in
generating the former event file(s).

We point out that data and event files may be very large. If the user
wants to store them in a scratch area, this can be done by setting the
script variable {\variable SCRTCH} equal to the physical address
of the scratch area (see sect.~\ref{sec:res}).

\subsection{Inclusive NLO cross sections\label{sec:xsecs}}
MC@NLO integrates NLO matrix elements in order to produce the
event file, and thus computes (as a by-product) the inclusive
NLO cross section. This cross section (whose value is given
in $pb$) can be obtained from an MC@NLO run in three different 
ways\footnote{This is true only if {\variable WGTTYPE}=1.}:
\begin{itemize}
\item[{\em a)}] It is printed out at the end of the NLO run (search for
{\tt Total for fully inclusive} in the standard output).
\item[{\em b)}] It is printed by HERWIG at the end of the MC run
(search for {\tt CROSS SECTION (PB)} in the standard output).
\item[{\em c)}] It is equal to the integral of any differential 
distribution which covers the whole kinematically-accessible range
(e.g. $0\le\pt\le\infty$) and on which no cuts are applied.
\end{itemize}
These three numbers are the same ({\em up to statistics}, which here
means the number of generated events -- see the bottom of this section
for further comments) for the processes listed in table~\ref{tab:proc}.
For the processes listed in table~\ref{tab:procdec}, on the other
hand, the results of {\em b)} and {\em c)} are equal to
that of {\em a)}, times the branching ratio(s) for the selected
decay channel(s), times (in the case of top decays) other factors 
due to kinematic cuts specified in input (see below). This is so 
because for the processes of table~\ref{tab:procdec} spin correlations 
are obtained as described in ref.~\cite{Frixione:2007zp}. For these
processes, we shall denote in what follows the cross section obtained
in {\em a)} as the undecayed cross section, and those obtained in
{\em b)} or {\em c)} as the decayed cross sections. We note that,
both for the processes in table~\ref{tab:proc} and for those in
table~\ref{tab:procdec}, the results of {\em b)} and {\em c)} are 
equal to the sum of the weights of all events stored in the
event file (possibly up to the contributions of those few events
which \HW\ is unable to shower and hadronize, and which are
therefore discarded with error messages in the MC run).

The branching ratios used in the computation are determined by the
values of the branching ratios for individual decay channels. The 
following variables are relevant to top decays:
\beq
{\variable BRTOPTOLEP}=\frac{\Gamma\left(\sum_j t\to l\nu_l b_j\right)}
                            {\Gamma_t}\,,
\;\;\;\;\;\;\;\;
{\variable BRTOPTOHAD}=\frac{\Gamma\left(\sum_{ij} t\to u \bar{d}_i b_j\right)}
                            {\Gamma_t}\,,
\label{eq:BRtop}
\eeq
with $b_j$ and $\bar{d}_i$ any down-type quark and antiquark respectively,
$u$ an up-type quark, and $l$ a charged lepton; lepton and flavour universality
are assumed. In the case of $W$ decays, one has the analogous variables
\beq
{\variable BRWTOLEP}=\frac{\Gamma\left(W\to l\nu_l\right)}
                            {\Gamma_W}\,,
\;\;\;\;\;\;\;\;
{\variable BRWTOHAD}=\frac{\Gamma\left(\sum_i W\to u \bar{d}_i\right)}
                            {\Gamma_W}\,.
\label{eq:BRW}
\eeq
The variables in eqs.~(\ref{eq:BRtop}) and~(\ref{eq:BRW}) can either be
given a numerical value in input, or computed at the LO in the SM by the
code -- see sect.~\ref{sec:decay} for details. The numerical values of
these variables are then combined to obtain the overall branching ratio
for the decay channels selected, which is done by setting the variables
{\variable IL}$_\alpha$ and {\variable TOPDECAY} as explained in 
sect.~\ref{sec:decay} (see in particular table~\ref{tab:ILval}).
For example, for a top decaying into a $W$ and any down-type 
quarks, with the $W$ decaying in an electron, muon, or any quarks, 
one sets {\variable IL}$_\alpha$=6, {\variable TOPDECAY=ALL},
and the resulting branching ratio will be 
$2\,\times\,${\variable BRTOPTOLEP}$+2\,\times\,${\variable BRTOPTOHAD}.

As mentioned above, in the case of top decays (as opposed to hard $W$ 
decays in $Wt$ or $W^+W^-$ production) the decayed cross section
will include kinematic factors in addition to the branching ratios.
These factors are due to the fact that in general the range for the invariant
mass of the pair of particles emerging from the $W$ decay (i.e. the 
virtuality of the $W$) does not coincide with the maximum that is 
kinematically allowed. For each top that decays, the following 
kinematic factor will therefore be included in the decayed cross
section
\beq
\frac{\Gamma\left(t\to ff^\prime b\,|\,
q_{\sss W}({\rm inf}),q_{\sss W}({\rm sup})\right)}
{\Gamma\left(t\to ff^\prime b\,|\,0,m_t\right)}\,,
\eeq
with
\beq
\Gamma\left(t\to ff^\prime b\,|\,m,M\right)=
\int_{m^2}^{M^2} dq_{\sss W}^2\frac{d\Gamma\left(t\to ff^\prime b\right)}
{dq_{\sss W}^2}\,,
\eeq
and $q_{\sss W}({\rm inf})$, $q_{\sss W}({\rm sup})$ the lower and 
upper limits of the $W$ virtuality, which can be chosen in input.
In particular, if {\variable V1GAMMAX}$>0$, one will have
\beq
q_{\sss W}({\rm inf})={\variable WMASS}-
{\variable V1GAMMAX}\,\times\,{\variable WWIDTH}\,,
\;\;\;\;\;\;\;\;
q_{\sss W}({\rm sup})={\variable WMASS}+
{\variable V1GAMMAX}\,\times\,{\variable WWIDTH}\,.
\label{eq:rangeA}
\eeq
On the other hand, if {\variable V1GAMMAX} $<0$, one has
\beq
q_{\sss W}({\rm inf})={\variable V1MASSINF}\,,
\;\;\;\;\;\;\;\;
q_{\sss W}({\rm sup})={\variable V1MASSSUP}\,.
\label{eq:rangeB}
\eeq
The ranges in eqs.~(\ref{eq:rangeA}) or~(\ref{eq:rangeB}) apply
to the $W$ emerging from the decay of the top quark in $t\bar{t}$ production,
and of the top or antitop in single-top production (all channels). The
corresponding ranges for the $W$ emerging from the decay of the antitop 
quark in $t\bar{t}$ production are identical to those above, except for
the replacement of {\variable V1} with {\variable V2}.

The user is also allowed to generate events by fixing the virtuality of the 
$W$ emerging from top/antitop decays equal to the $W$ pole mass, by
setting {\variable xGAMMAX}$=0$, with {\variable x=V1,V2}. In such a
case, the decayed cross section will be equal to the undecayed cross
section, times the branching ratios, times a factor
\beq
\frac{d\Gamma\left(t\to ff^\prime b\right)}{dq_{\sss W}^2}
\Bigg|_{q_{\sss W}^2=M_{\sss W}^2}\,,
\eeq
for each decaying top quark. The decayed cross section will have 
therefore to be interpreted as differential in the $W$ virtuality squared
(doubly differential in the case of $t\bar{t}$ production), and
will be expressed in \mbox{$pb~{\rm GeV}^{-2}$} 
(or \mbox{$pb~{\rm GeV}^{-4}$} for $t\bar{t}$ production) units.

The branching ratios and kinematics factors for each decaying particles
are multiplied to give a single number (always less than or equal
to one), which is by definition the ratio of the decayed over the
undecayed cross section. This number is printed out at the end of the 
NLO run (search for {\tt Normalization factor due to decays} in the 
standard output).

We conclude this section by stressing that, 
while the result of {\em a)} is always computed with
a typical relative precision of $10^{-4}$, those of {\em b)} and
{\em c)} depend on the number of events generated. Although it has
been checked that, upon increasing the number of events generated, the
results of {\em b)} and {\em c)} do approach that of {\em a)} (possibly
times the branching ratios and kinematic factors), option {\em a)} has 
clearly to be preferred. As mentioned above, the decayed 
cross section of {\em b)} or {\em c)} can be obtained without
any loss of accuracy by multiplying the undecayed cross section of
{\em a)} by the normalization factor printed out by the code at
the end of the NLO run.

\subsection{$Wt$ production}\label{sec:Wt}
Owing to the interference with $t\bar{t}$ production, which occurs in 
the $gg$ and $q\bar{q}$ partonic channels starting at the NLO, the $Wt$ 
cross section is ill-defined beyond the leading order in QCD. One can 
still give an operative meaning to NLO $Wt$ production, but one must
always be aware of the potential biases introduced in this way.
This issue and its potential physics implications are discussed
at length in ref.~\cite{Frixione:2008yi}, which the reader is
strongly advised to consult before generating $Wt$ events.

In MC@NLO version 3.4, we have implemented two different definitions
of the $Wt$ cross section, which we denoted by {\em diagram removal}
and {\em diagram subtraction} in ref.~\cite{Frixione:2008yi}.
The former computation is carried out by setting {\variable WTTYPE=REMOVAL}
in {\code MCatNLO.inputs}, while the latter corresponds to
{\variable WTTYPE=SUBTRACTION}. 

In $Wt$ production, the factorization (renormalization) scale is 
assigned the value of the variable {\variable PTVETO} (whose units
are GeV) if {\variable FFACT}$<0$ ({\variable FREN}$<0$). This option
should be used for testing purposes only; it is not recommended
in the generation of event samples for experimental studies.

\subsection{Decays}\label{sec:decay}
$\MCatNLO$ is intended primarily for the study of NLO corrections
to production cross sections and distributions; NLO corrections to
the decays of produced particles are not included. As for spin 
correlations, the situation in version 3.4 is summarized 
in tables~\ref{tab:proc} and~\ref{tab:procdec}: they are included for all 
processes except $ZZ$ and $WZ$ production\footnote{Non-factorizable 
spin correlations of virtual origin are not included in $W^+W^-$,
$t\bar{t}$, and single-$t$ production. See ref.~\cite{Frixione:2007zp}.}.
For the latter processes, quantities sensitive to the polarisation of 
produced particles are not given correctly even to leading order.
For such quantities, it may be preferable to use the standard
\HW\ MC, which does include leading-order spin correlations.

Following \HW\ conventions, spin correlations in single-vector-boson
processes are automatically included using the process codes
({\variable IPROC}) relevant to lepton pair production (in other
words, if one is interested in including spin correlations in e.g. 
$W^+$ production and subsequent decays into $\mu^+\nu_\mu$, one needs to use
{\variable IPROC}$\,=\!-1461$ rather than {\variable IPROC}$\,=\!-1497$
and {\variable MODBOS}$(1)=3$). In order to avoid an 
unnecessary proliferation of {\variable IPROC} values, this strategy 
has not been adopted in other cases ($t\bar{t}$, single-$t$, $H^0W^\pm$, 
$H^0Z$, $W^+W^-$), in which spin correlations are included if
the variables {\variable IL}$_1$ and {\variable IL}$_2$ (the
latter is used only in $t\bar{t}$, $Wt$, and $W^+W^-$ production) are 
assigned certain values. In the case of individual lepton decays,
these range from 1 to 3 if the decaying particle is a $W$ or a top,
or from 1 to 6 if the decaying particle is a $Z$. For these cases, the value of
{\variable IL}$_\alpha$ fully determines the identity of the leptons emerging
from the decay, and the same convention as in \HW\ is adopted
(see the \HW\ manual and sect.~\ref{sec:oper}).

In $t\bar{t}$ and single-top production, i.e. for all processes
listed in table~\ref{tab:procdec}, the top quark and/or antiquark,
and the hard $W$ in the case of $Wt$ production, can also decay hadronically.
In such cases, therefore, the variables {\variable IL}$_\alpha$ can
be assigned more values than for the other processes; the situation
is summarized in table~\ref{tab:ILval}. When generating the decays,
lepton and flavour universalities are assumed. The relative probabilities
of individual hadronic decays (e.g. $W^+\to u\bar{d}$ vs $W^+\to u\bar{s}$)
are determined using the CKM matrix elements entered by the user
(variables {\variable Vud} in {\code MCatNLO.inputs}). The relative
probabilities of leptonic vs hadronic decays are on the other hand
determined using the values of the corresponding branching ratios
entered by the user: variables {\variable BRTOPTOLEP} and
{\variable BRTOPTOHAD} for top/antitop decays, and {\variable BRWTOLEP}
and {\variable BRWTOHAD} for the decays of the hard $W$ emerging from
the hard process in $Wt$ production\footnote{{\variable BRWTOLEP} is
also used in $W^+W^-$ production. $W$ hadronic decays are not implemented
in this process, hence the branching ratio is only used as a rescaling
factor for event weights.} -- see eqs.~(\ref{eq:BRtop}) and~(\ref{eq:BRW})
for the definitions of these variables.
\begin{table}[htb]
\begin{center}
\begin{tabular}{|c|c|}\hline
{\variable IL}$_\alpha$ & Decay \\
\hline\hline
0 & $e+\mu+\tau+q$ \\\hline
1 & $e$ \\\hline
2 & $\mu$ \\\hline
3 & $\tau$ \\\hline
4 & $e+\mu$ \\\hline
5 & $q$ \\\hline
6 & $e+\mu+q$ \\\hline
7 & no decay \\\hline
\end{tabular}
\end{center}
\caption{\label{tab:ILval}
Decays of the $W$'s originating from top/antitop decay or from the 
hard process in $Wt$ production. The symbol $q$ denotes all hadronic
$W$ decays. Values different from 1, 2, or 3 are only allowed in
$t\bar{t}$ and single-top production (all channels).}
\end{table}

In the case of top/antitop decays, it is also possible to generate events 
in which the top decays into a $W$ and any down-type quark (hence the 
notations $b_\alpha$ and $\bar{b}_\alpha$ in table~\ref{tab:procdec}).
The identity of the latter is determined according to the CKM matrix 
values. For this to happen, one needs to set {\variable TOPDECAY=ALL}
in {\code MCatNLO.inputs}. If, on the other hand, one wants to
always generate $t\to Wb$ decays, one needs to set 
{\variable TOPDECAY=Wb}; in such a case, event weights (and thus
the decayed cross section, as defined in sect.~\ref{sec:xsecs})
will be multiplied by a factor $V_{tb}^2/(V_{td}^2+V_{ts}^2+V_{tb}^2)$.

For the processes in table~\ref{tab:procdec} it is also possible to 
force the code to use the LO values of the relevant leptonic and
hadronic branching ratios, by entering negative values for the
top and $W$ widths (variables {\variable TWIDTH} and
{\variable WWIDTH} in {\code MCatNLO.inputs}). In such a case,
the values of {\variable BRTOPTOLEP}, {\variable BRTOPTOHAD}, 
{\variable BRWTOLEP} and {\variable BRWTOHAD} given in the
input file will be ignored, and replaced by $1/9$, $1/3$, $1/9$
and $1/3$ respectively. The top and $W$ widths will be computed
using the LO SM formulae.

Spin correlations are implemented in the processes in table~\ref{tab:procdec} 
according to the method of ref.~\cite{Frixione:2007zp}, which is based
on a zero-width approximation for the decaying particles. Nevertheless,
the top quark and antiquark in $t\bar{t}$ production 
({\variable IPROC}$\,=\!-1706$), and the vector bosons in $W^+W^-$
production ({\variable IPROC}$\,=\!-2850$) can be given masses 
different from the pole masses. These off-shell effects are modeled
by re-weighting the cross section with skewed Breit-Wigner functions
(in order to take into account the fact that by changing the invariant
mass of the system produced one probes different values of Bjorken
$x$'s). This re-weighting is unitary, i.e. it does not change the
inclusive cross section. For $t\bar{t}$ production, the ranges
of top and antitop masses are controlled by the parameters
{\variable TiGAMMAX}, {\variable TiMASSINF}, and {\variable TiMASSSUP}
(with {\variable i}=1,2 for top and antitop respectively).
For $W^+W^-$, one needs to use instead {\variable ViGAMMAX}, 
{\variable ViMASSINF}, and {\variable ViMASSSUP}, with
{\variable i}=1,2 for $W^+$ and $W^-$ respectively. In both cases,
the mass ranges will be defined by formulae formally identical to
those of eqs.~(\ref{eq:rangeA}) and~(\ref{eq:rangeB}).
In version 3.4, off-shell effects are not implemented in the other
processes in table~\ref{tab:procdec}, i.e. all channels of
single-top production.

Finally, we point out that since spin correlations for the processes 
in table~\ref{tab:procdec} are implemented according to the method
of ref.~\cite{Frixione:2007zp}, tree-level matrix elements for
leptonic final states are needed. The codes for these have been generated
with MadGraph/MadEvent~\cite{Stelzer:1994ta,Maltoni:2002qb}, and
embedded into the MC@NLO package.

When {\variable IL}$_\alpha$=7, the corresponding particle is left undecayed 
by the \NLO\ code, and is passed as such to the \MC\ code; the information
on spin correlations is lost. However, the user can still force particular 
decay modes during the \MC\ run. In the case of vector bosons, one proceeds
in the same way as in standard \HW, using the {\variable MODBOS}
variables -- see sect.~3.4 of ref.~\cite{Corcella:2001bw}. However,
top decays cannot be forced in this way because the decay is
treated as a three-body process: the $W^\pm$ boson entry in
{\code HEPEVT} is for information only.  Instead, the top
branching ratios can be altered using the {\variable HWMODK}
subroutine -- see sect.~7 of ref.~\cite{Corcella:2001bw}.
This is done separately for the $t$ and $\bar t$. For example,
{\code CALL HWMODK(6,1.D0,100,12,-11,5,0,0)} forces
the decay $t\to \nu_e e^+ b$, while leaving $\bar t$ decays
unaffected.  Note that the order of the decay products is
important for the decay matrix element
({\variable NME} = 100) to be applied correctly.
The relevant statements should be inserted in the \HW\ main program
(corresponding to {\code mcatnlo\_hwdriver.f} in this package)
after the statement {\code CALL HWUINC} and before
the loop over events.  A separate run with
{\code CALL HWMODK(-6,1.D0,100,-12,11,-5,0,0)} should
be performed if one wishes to symmetrize the forcing of
$t$ and $\bar t$ decays, since calls to {\variable HWMODK} from
within the event loop do not produce the desired result.

\subsection{Results\label{sec:res}}
As in the case of standard \HW\, the form of the results will be
determined by the user's analysis routines. However, in addition
to any files written by the user's analysis routines, the
$\MCatNLO$ writes the following files:\\
$\blacklozenge$ 
{\variable FPREFIXNLOinput}: the input file for the \NLO\ executable, 
created according to the set of input parameters defined in 
{\code MCatNLO.inputs} (where the user also sets the string
{\variable FPREFIX}). See table~\ref{tab:NLOi}.\\
$\blacklozenge$ 
{\variable FPREFIXNLO.log}: the log file relevant to the \NLO\ run.\\
$\blacklozenge$ 
{\variable FPREFIXxxx.data}: {\variable xxx} can assume several different 
values. These are the data files created by the \NLO\ code. They can be 
removed only if no further event generation step is foreseen with the
current choice of parameters.\\
$\blacklozenge$ 
{\variable FPREFIXMCinput}: analogous to {\variable FPREFIXNLOinput}, 
but for the \MC\ executable. See table~\ref{tab:MCi}.\\
$\blacklozenge$ 
{\variable FPREFIXMC.log}: analogous to {\variable FPREFIXNLO.log}, but 
for the \MC\ run.\\
$\blacklozenge$ 
{\variable EVPREFIX.events}: the event file, where {\variable EVPREFIX} 
is the string set by the user in {\code MCatNLO.inputs}.\\
$\blacklozenge$ 
{\variable EVPREFIXxxx.events}: {\variable xxx} can assume several different 
values. These files are temporary event files, which are used by the
\NLO\ code, and eventually removed by the shell scripts. They MUST NOT be
removed by the user during the run (the program will crash or give
meaningless results).

By default, all the files produced by the $\MCatNLO$ are written in the
running directory.  However, if the variable {\variable SCRTCH} (to be set in
{\code MCatNLO.inputs}) is {\em not} blank, the data and event files will be
written in the directory whose address is stored in {\variable SCRTCH}
(such a directory is not created by the scripts, and must already exist
at run time).

\section{Script variables\label{sec:scrvar}}
In the following, we list all the variables appearing in 
{\code MCatNLO.inputs}; these can be changed by the user to suit 
his/her needs. This must be done by editing {\code MCatNLO.inputs}.
For fuller details see the comments in {\code MCatNLO.inputs}.
\begin{itemize}
\item[{\variable ECM}] 
 The CM energy (in GeV) of the colliding particles.
\item[{\variable FREN}] 
 The ratio between the renormalization scale, and a reference mass scale.
\item[{\variable FFACT}] 
 As {\variable FREN}, for the factorization scale.
\item[{\variable HVQMASS}] 
 The mass (in GeV) of the top quark, except when 
 {\variable IPROC}$\,=\!-(1)1705$,
 when it is the mass of the bottom quark. In this case, {\variable HVQMASS} 
 must coincide with {\variable BMASS}.
\item[{\variable xMASS}] 
 The mass (in GeV) of the particle {\variable x}, with 
 {\variable x=HGG,W,Z,U,D,S,C,B,G}.
\item[{\variable xWIDTH}] 
 The physical (Breit-Wigner) width (in GeV) of the particle {\variable x}, 
 with {\variable x=HGG,W,Z,T} for $H^0$, $W^\pm$, $Z$, and $t$ respectively.
\item[{\variable BRTOPTOx}] 
 Branching ratio for top decay channels $\sum_j t\to l\nu_l b_j$ (when
 {\variable x=LEP}) and $\sum_{ij} t\to u \bar{d}_i b_j$ (when
 {\variable x=HAD}). Lepton and flavour universality is assumed.
\item[{\variable BRWTOx}] 
 Branching ratio for $W$ decay channels $W\to l\nu_l$ (when
 {\variable x=LEP}) and $\sum_i W\to u \bar{d}_i$ (when
 {\variable x=HAD}). Lepton and flavour universality is assumed.
\item[{\variable IBORNHGG}] 
 Valid entries are 1 and 2.  If set to 1, the exact top mass dependence is
 retained {\em at the Born level} in Higgs production.  If set to 2, the
 $m_t\to\infty$ limit is used.
\item[{\variable xGAMMAX}] 
 If {\variable xGAMMAX} $>0$, controls the width of the mass range for 
 Higgs ({\variable x=H}), vector bosons ({\variable x=V1,V2}), and
 top ({\variable x=T1,T2}): the range is 
 ${\variable MASS}\pm({\variable GAMMAX} \times{\variable WIDTH})$.
 Off-shell effects for top are only implemented in $t\bar{t}$ production.
\item[{\variable xMASSINF}] 
 Lower limit of the Higgs ({\variable x=H}), vector boson
 ({\variable x=V1,V2}), and top ({\variable x=T1,T2})
 mass range; used only when {\variable xGAMMAX} $<0$.
\item[{\variable xMASSSUP}] 
 Upper limit of the Higgs ({\variable x=H}), vector boson 
 ({\variable x=V1,V2}), and top ({\variable x=T1,T2})
 mass range; used only when {\variable xGAMMAX} $<0$.
\item[{\variable Vud}]
 CKM matrix elements, with {\variable u}={\variable U,C,T} and
 {\variable d}={\variable D,S,B}. Set {\variable VUD=VUS=VUB}=0
 to use values of PDG2003. 
\item[{\variable AEMRUN}]
 Set it to {\variable YES} to use running $\alpha_{em}$ in lepton pair and
 single vector boson production, set it to {\variable NO} to use 
 $\alpha_{em}=1/137.0359895$.
\item[{\variable IPROC}]
 Process number that identifies the hard subprocess: see tables~\ref{tab:proc} 
 and~\ref{tab:procdec} for valid entries.
\item[{\variable IVCODE}]
 Identifies the nature of the vector boson in associated Higgs production.
 It corresponds to variable {\variable IV} of table~\ref{tab:proc}.
\item[{\variable ILxCODE}]
 Identify the nature of the particles emerging from vector boson or top 
 decays. They correspond to variables {\variable IL}$_1$ and 
 {\variable IL}$_2$ (for {\variable x} $=1,2$ respectively) of 
 tables~\ref{tab:proc}, \ref{tab:procdec} and~\ref{tab:ILval}.
\item[{\variable TOPDECAY}]
 Valid entries are {\variable ALL} and {\variable Wb}. Controls the type
 of top decay. See sect.~\ref{sec:decay}.
\item[{\variable WTTYPE}]
 Valid entries are {\variable REMOVAL} and {\variable SUBTRACTION}. Determines
 the definition of the $Wt$ cross section at the NLO. See sect.~\ref{sec:Wt}.
\item[{\variable PTVETO}]
 Used in conjunction with {\variable FFACT} and/or {\variable FREN} to 
 set mass scales in $Wt$ production. See sect.~\ref{sec:Wt}.
\item[{\variable PARTn}]
 The type of the incoming particle \#{\variable n}, with {\variable n}=1,2. 
 \HW\ naming conventions are used ({\variable P, PBAR, N, NBAR}).
\item[{\variable PDFGROUP}]
 The name of the group fitting the parton densities used;
 the labeling conventions of PDFLIB are adopted. Unused when linked
 to LHAPDF.
\item[{\variable PDFSET}] 
 The number of the parton density set; according to PFDLIB conventions,
 the pair ({\variable PDFGROUP}, {\variable PDFSET}) identifies the 
 densities for a given particle type. When linked to LHAPDF, use 
 the numbering conventions of LHAGLUE~\cite{Whalley:2005nh}.
\item[{\variable LAMBDAFIVE}]
 The value of $\Lambda_{\sss QCD}$, for five flavours and in the 
 ${\overline {\rm MS}}$ scheme, used in the computation of NLO
 cross sections. A negative entry sets $\Lambda_{\sss QCD}$ equal
 to that associated with the PDF set being used.
\item[{\variable LAMBDAHERW}]
 The value of $\Lambda_{\sss QCD}$ used in MC runs; this parameter has the 
 same meaning as $\Lambda_{\sss QCD}$ in \HW.
\item[{\variable SCHEMEOFPDF}] 
 The subtraction scheme in which the parton densities are defined.
\item[{\variable FPREFIX}] Our integration routine creates files with
 name beginning by the string {\variable FPREFIX}. Most of these files are not 
 directly accessed by the user. See sects.~\ref{sec:evfile} 
 and~\ref{sec:res}.
\item[{\variable EVPREFIX}] 
 The name of the event file begins with this string. See 
 sects.~\ref{sec:evfile} and~\ref{sec:res}.
\item[{\variable EXEPREFIX}] 
 The names of the \NLO\ and \MC\ executables begin with this string; this is
 useful in the case of simultaneous runs.
\item[{\variable NEVENTS}] 
 The number of events stored in the event file, eventually
 processed by \mbox{\HW\ .}
\item[{\variable WGTTYPE}]
 Valid entries are 0 and 1. When set to 0, the weights in the event file
 are $\pm 1$. When set to 1, they are $\pm w$, with $w$ a constant such
 that the sum of the weights gives the total inclusive NLO cross section
 (see sect.~\ref{sec:xsecs} for more details).
 Note that these weights are redefined by \HW\ at \MC\ run time according 
 to its own convention (see \HW\ manual).
\item[{\variable RNDEVSEED}] 
 The seed for the random number generation in the
 event generation step; must be changed in order to obtain
 statistically-equivalent but different event files.
\item[{\variable BASES}] 
 Controls the integration step; valid entries are {\variable ON} and 
 {\variable OFF}. At least one run with {\variable BASES=ON} must be 
 performed (see sect.~\ref{sec:evfile}).
\item[{\variable PDFLIBRARY}] 
 Valid entries are {\variable PDFLIB}, {\variable LHAPDF}, and 
 {\variable THISLIB}. In the former two cases, PDFLIB or LHAPDF is used to 
 compute the parton densities, whereas in the latter case the densities are
 obtained from our self-contained PDF library.
\item[{\variable HERPDF}] 
 If set to {\variable DEFAULT}, \HW\ uses its internal PDF set 
 (controlled by {\variable NSTRU}), regardless of the densities
 adopted at the NLO level. If set to {\variable EXTPDF}, \HW\ uses
 the same PDFs as the \NLO\ code (see sect.~\ref{sec:pdfs}).
\item[{\variable HWPATH}]
 The physical address of the directory where the user's
 preferred version of \HW\ is stored.
\item[{\variable SCRTCH}]
 The physical address of the directory where the user wants to store the
 data and event files. If left blank, these files are stored in the 
 running directory.
\item[{\variable HWUTI}]
This variables must be set equal to a list of object files,
needed by the analysis routines of the user (for example,
{\variable HWUTI=''obj1.o obj2.o obj3.o''} is a valid assignment).
\item[{\variable HERWIGVER}]
This variable must to be set equal to the name of the
object file corresponding to the version of \HW\ linked
to the package (for example, {\variable HERWIGVER=herwig6510.o} is a
valid assignment).
\item[{\variable PDFPATH}]
 The physical address of the directory where the PDF grids are stored.
 Effective only if {\variable PDFLIBRARY=THISLIB}.
\item[{\variable LHAPATH}]
 Set this variable equal to the name of the directory where the local 
 version of LHAPDF is installed. See sect.~\ref{sec:lhapdf}.
\item[{\variable LHAOFL}]
 Set {\variable LHAOFL=FREEZE} to freeze PDFs from LHAPDF at the boundaries,
 or equal to {\variable EXTRAPOLATE} otherwise. See LHAPDF manual for
 details.
\item[{\variable EXTRALIBS}]
 Set this variable equal to the names of the libraries which need be linked.
 LHAPDF is a special case, and must not be included in this list.
\item[{\variable EXTRAPATHS}]
 Set this variable equal to the names of the directories where the 
 libraries which need be linked are installed.
\item[{\variable INCLUDEPATHS}]
 Set this variable equal to the names of the directories which 
 contain header files possibly needed by C++ files provided by the user
 (via {\variable HWUTI}). 
\end{itemize}

\section*{Acknowledgments}
It is a pleasure to thank the co-authors of the MC@NLO papers,
E.~Laenen, P.~Motylinski, P.~Nason, and C.~D.~White, for having
contributed so much to many different aspects of the $\MCatNLO$ project,
and for stimulating discussions. We thank W.~Verkerke for having 
provided us with a Fortran interface to C++ Root-calling routines.
BRW thanks the CERN theory group for frequent hospitality.
Finally, we are indebted with all the members of experimental 
collaborations, unfortunately too numerous to be explicitly mentioned 
here, who used the code and gave us precious suggestions and feedback.

\section*{Appendices}
\appendix

\section{Version changes}

\subsection{From MC@NLO version 1.0 to version 2.0\label{app:newver}}
In this appendix we list the changes that occurred in the package
from version 1.0 to version 2.0.

$\bullet$~The Les Houches generic user process interface has been adopted.

$\bullet$~As a result, the convention for process codes has been changed:
MC@NLO process codes {\variable IPROC} are negative.

$\bullet$~The code {\code mcatnlo\_hwhvvj.f}, which was specific to
vector boson pair production in version 1.0, has been replaced by
{\code mcatnlo\_hwlhin.f}, which reads the event file according
to the Les Houches prescription, and works for all the production 
processes implemented.

$\bullet$~The {\code Makefile} need not be edited, since the variables
{\variable HERWIGVER} and {\variable HWUTI} have been moved to
{\code MCatNLO.inputs} (where they must be set by the user).

$\bullet$~A code {\code mcatnlo\_hbook.f} has been added to the list of
utility codes. It contains a simplified version (written by M.~Mangano)
of {\small HBOOK}, and it is only used by the sample analysis routines
{\code mcatnlo\_hwan{\em xxx}.f}. As such, the user will not need it
when linking to a self-contained analysis code.

We also remind the reader that the \HW\ version must be 
6.5 or higher since the  Les Houches interface is used.

\subsection{From MC@NLO version 2.0 to version 2.1\label{app:newvera}}
In this appendix we list the changes that occurred in the package
from version 2.0 to version 2.1.

$\bullet$~Higgs production has been added, which implies new process-specific 
files\\ ({\code mcatnlo\_hgmain.f}, {\code mcatnlo\_hgxsec.f}, 
{\code hgscblks.h}, {\code mcatnlo\_hwanhgg.f}), and a modification to
{\code mcatnlo\_hwlhin.f}. 

$\bullet$~Post-1999 PDF sets have been added to the MC@NLO PDF library.

$\bullet$~Script variables have been added to {\code MCatNLO.inputs}. 
Most of them are only relevant to Higgs production, and don't affect processes
implemented in version 2.0. One of them ({\variable LAMBDAHERW}) may affect
all processes: in version 2.1, the variables {\variable LAMBDAFIVE} and
{\variable LAMBDAHERW} are used to set the value of $\Lambda_{\sss QCD}$ in
NLO and MC runs respectively, whereas in version 2.0 {\variable LAMBDAFIVE}
controlled both. The new setup is necessary since modern PDF sets have
$\Lambda_{\sss QCD}$ values which are too large to be supported by \HW.
(Recall that the effect of using {\variable LAMBDAHERW} different from
{\variable LAMBDAFIVE} is beyond NLO.)

$\bullet$~The new script variable {\variable PDFPATH} should be set equal 
to the name of the directory where the PDF grid files (which can be downloaded
from the MC@NLO web page) are stored. At run time, when executing {\code
runNLO}, or {\code runMC}, or {\code runMCatNLO}, logical links to these
files will be created in the running directory (in version 2.0, this
operation had to be performed by the user manually).

$\bullet$~Minor bugs corrected in {\code mcatnlo\_hbook.f} and sample 
analysis routines.

\subsection{From MC@NLO version 2.1 to version 2.2\label{app:newverb}}
In this appendix we list the changes that occurred in the package
from version 2.1 to version 2.2.

$\bullet$~Single vector boson production has been added, which implies 
new process-specific files ({\code mcatnlo\_sbmain.f}, 
{\code mcatnlo\_sbxsec.f}, {\code svbcblks.h}, {\code mcatnlo\_hwansvb.f}), 
and a modification to {\code mcatnlo\_hwlhin.f}. 

$\bullet$~The script variables {\variable WWIDTH} and {\variable ZWIDTH}
have been added to {\code MCatNLO.inputs}. These denote the physical widths 
of the $W$ and $Z^0$ bosons, used to generate the mass distributions of
the vector bosons according to the Breit--Wigner function, in the case
of single vector boson production (vector boson pair production is
still implemented only in the zero-width approximation).

\subsection{From MC@NLO version 2.2 to version 2.3\label{app:newverc}}
In this appendix we list the changes that occurred in the package
from version 2.2 to version 2.3.

$\bullet$~Lepton pair production has been added, which implies 
new process-specific files ({\code mcatnlo\_llmain.f}, 
{\code mcatnlo\_llxsec.f}, {\code llpcblks.h}, {\code mcatnlo\_hwanllp.f}), 
and modifications to {\code mcatnlo\_hwlhin.f} and 
{\code mcatnlo\_hwdriver.f}.

$\bullet$~The script variable {\variable AEMRUN} has been added, since
the computation of single vector boson and lepton pair cross sections is
performed in the $\MSbar$ scheme (the on-shell scheme was previously
used for single vector boson production).

$\bullet$~The script variables {\variable FRENMC} and {\variable FFACTMC}
have been eliminated. 

$\bullet$~The structure of pseudo-random number generation in heavy
flavour production has been changed, to avoid a correlation that
affected the azimuthal angle distribution for the products of the hard
partonic subprocesses.

$\bullet$~A few minor bugs have been corrected, which affected the rapidity
of the vector bosons in single vector boson production (a 2--3\% effect),
and the assignment of $\Lambda_{\sss QCD}$ for the LO and NLO PDF sets of
Alekhin.

\subsection{From MC@NLO version 2.3 to version 3.1\label{app:newverd}}
In this appendix we list the changes that occurred in the package
from version 2.3 to version 3.1.

$\bullet$~Associated Higgs production has been added, which implies 
new process-specific files ({\code mcatnlo\_vhmain.f}, 
{\code mcatnlo\_vhxsec.f}, {\code vhgcblks.h}, {\code mcatnlo\_hwanvhg.f}), 
and modifications to {\code mcatnlo\_hwlhin.f} and 
{\code mcatnlo\_hwdriver.f}.

$\bullet$~Spin correlations in $W^+W^-$ production and leptonic decay
have been added;
the relevant codes ({\code mcatnlo\_vpmain.f}, {\code mcatnlo\_vhxsec.f})
have been modified; the sample analysis routines ({\code mcatnlo\_hwanvbp.f})
have also been changed. Tree-level matrix elements have been computed with
MadGraph/MadEvent~\cite{Stelzer:1994ta,Maltoni:2002qb}, which uses 
HELAS~\cite{Murayama:1992gi}; the relevant routines and common blocks 
are included in {\code mcatnlo\_helas2.f} and {\code MEcoupl.inc}.

$\bullet$~The format of the event file has changed in several respects,
the most relevant of which is that the four-momenta are now given as
$(p_x,p_y,p_z,m)$ (up to version 2.3 we had $(p_x,p_y,p_z,E)$). Event
files generated with version 2.3 or lower {\em must not be used} with
version 3.1 or higher (the code will prevent the user from doing so).

$\bullet$~The script variables {\variable GAMMAX}, {\variable MASSINF},
and {\variable MASSSUP} have been replaced with {\variable xGAMMAX}, 
{\variable xMASSINF} and {\variable xMASSSUP}, with {\variable x=H,V1,V2}.

$\bullet$~New script variables {\variable IVCODE}, {\variable IL1CODE},
and {\variable IL2CODE} have been introduced.

$\bullet$~Minor changes have been made to the routines that put the partons
on the \HW\ mass shell for lepton pair, heavy quark, and vector boson pair
production; effects are beyond the fourth digit.

$\bullet$~The default electroweak parameters have been changed for
vector boson pair production, in order to make them consistent with those
used in other processes. The cross sections are generally smaller
in version 3.1 wrt previous versions, the dominant effect being the 
value of $\sinthW$: we have now $\sinsqthW=0.2311$, in lower
versions $\sinsqthW=1-m_W^2/m_Z^2$. The cross sections
are inversely proportional to $\sinfthW$.

\subsection{From MC@NLO version 3.1 to version 3.2\label{app:newvere}}
In this appendix we list the changes that occurred in the package
from version 3.1 to version 3.2.

$\bullet$~Single-$t$ production has been added, which implies 
new process-specific files\\ ({\code mcatnlo\_stmain.f}, 
{\code mcatnlo\_stxsec.f}, {\code stpcblks.h}, {\code mcatnlo\_hwanstp.f}), 
and modifications to {\code mcatnlo\_hwlhin.f} and 
{\code mcatnlo\_hwdriver.f}.

$\bullet$~LHAPDF library is now supported, which implies modifications to
all {\code *main.f} files, and two new utility codes,
{\code mcatnlo\_lhauti.f} and {\code mcatnlo\_mlmtolha.f}.

$\bullet$~New script variables {\variable Vud}, {\variable LHAPATH},
and {\variable LHAOFL} have been introduced.

$\bullet$~A bug affecting Higgs production has been fixed, which implies
a modification to {\code mcatnlo\_hgxsec.f}. Cross sections change with
respect to version 3.1 {\em only if} {\variable FFACT}$\ne 1$ (by 
${\cal O}(1\%)$ in the range $1/2\le$ {\variable FFACT} $\le 2$).

\subsection{From MC@NLO version 3.2 to version 3.3\label{app:newverf}}
In this appendix we list the changes that occurred in the package
from version 3.2 to version 3.3.

$\bullet$~Spin correlations have been added to $t\bar{t}$ and single-$t$
production processes, which imply modifications to several codes
({\code mcatnlo\_qqmain.f}, {\code mcatnlo\_qqxsec.f},
{\code mcatnlo\_stmain.f}, {\code mcatnlo\_stxsec.f},
{\code mcatnlo\_hwlhin.f} and {\code mcatnlo\_hwdriver.f}).
Tree-level matrix elements have been computed with
MadGraph/MadEvent~\cite{Stelzer:1994ta,Maltoni:2002qb}.

$\bullet$~The matching between NLO matrix elements and parton shower
is now smoother in Higgs production, which helps eliminate one unphysical 
feature in the $\pt$ spectra of the accompanying jets. The code 
{\code mcatnlo\_hgmain.f} has been modified. Technical details on this
matching procedure will be posted on the MC@NLO web page.

$\bullet$~The new script variable {\variable TWIDTH} has been introduce.d

$\bullet$~All instances of {\variable HWWARN('s',i,*n)} have been
replaced with {\variable HWWARN('s',i)} in \HW-related codes. This
is consistent with the definition of {\variable HWWARN} in \HW\ versions
6.510 and higher; the user must be careful if linking to \HW\ versions,
in which the former form of {\variable HWWARN} is used. Although \HW\ 6.510 
compiles with {\variable g95} or {\variable gfortran}, MC@NLO 3.3
does not.

\subsection{From MC@NLO version 3.3 to version 3.4\label{app:newverg}}
In this appendix we list the changes that occurred in the package
from version 3.3 to version 3.4.

$\bullet$~$Wt$ production has been implemented, which implies new 
process-specific codes ({\code mcatnlo\_wtmain\_dr.f},
{\code mcatnlo\_wtmain\_ds.f}, {\code mcatnlo\_wtxsec\_dr.f}\\ and
{\code mcatnlo\_wtxsec\_ds.f}).

$\bullet$~Owing to the implementation of $Wt$ production and of top
hadronic decays, the Les Houches interface ({\code mcatnlo\_hwlhin.f}) 
and the driver ({\code mcatnlo\_hwdriver.f}) have been upgraded.

$\bullet$~New script variables ({\variable BRTOPTOx} and 
{\variable BRWTOx}, with {\variable x=LEP,HAD}; {\variable yGAMMAX},
{\variable yMASSINF} and {\variable yMASSSUP} with {\variable y=T1,T2};
{\variable TOPDECAY}; {\variable WTTYPE}; {\variable PTVETO})
have been introduced.

$\bullet$~The new script variables {\variable EXTRALIBS}, 
{\variable EXTRAPATHS}, and {\variable INCLUDEPATHS} can be used to 
link to external libraries. Their use has only been tested on a recent 
Scientific Linux release, and they may be not portable to other systems.

$\bullet$~The ranges of variables {\variable ILxCODE} have been extended
for several processes, in order to account for the newly-implemented
hadronic decays.

$\bullet$~{\code MCatNLO.inputs} and {\code MCatNLO.Script} have been
upgraded to reflect the changes above. A new sample input file
({\code MCatNLO\_rb.inputs}) is included, which documents the
use of an analysis producing plots in Root format. Finally, the possibility
is given to link to a dynamic LHAPDF library (through 
{\code MCatNLO\_dyn.Script} and {\code Makefile\_dyn}).

$\bullet$~Front-end Fortran routines ({\code rbook\_fe.f}) are provided,
to produce plots in Root format, using the same syntax as for calling our
HBOOK-type routines. A companion C++ code is needed ({\code rbook\_be.cc}).
These codes have been written by W.~Verkerke. Examples of 
analysis routines using Root format have been added 
({\code mcatnlo\_hwan{\em xxx}\_rb.f}). A call to a release-memory
routine ({\code RCLOS}) has been added to {\code mcatnlo\_hwdriver.f}; 
this is only needed when using a Root-format output, and a dummy 
body of {\code RCLOS} has been added to HBOOK-format analysis
files {\code mcatnlo\_hwan{\em xxx}.f}.

$\bullet$~The linking to LHAPDF has been upgraded, assuming the use
of LHAPDF version 5.0 or higher. The file {\code mcatnlo\_lhauti.f}
has been eliminated, and replaced with {\code mcatnlo\_utilhav4.f},
which is however necessary only if the user wants to link with
LHAPDF versions 4.xx (in such a case, the user will also need to
edit the Makefile).

$\bullet$~The automatic assignment of $\Lambda_{\sss QCD}$ when
using LHAPDF is now to be considered robust. This implies changes
to {\code mcatnlo\_mlmtolha.f}, the insertion of a dummy routine
into {\code mcatnlo\_mlmtopdf.f} and {\code mcatnlo\_pdftomlm.f},
and very minor changes to all {\code *main*.f} files.

$\bullet$~Minor changes to {\code mcatnlo\_hbook.f}, mainly affecting
two-dimensional plot outputs.

$\bullet$~A bug has been fixed, which prevented one from choosing properly 
the $W$ mass ranges in $W^+W^-$ production and subsequent decays in
the case of {\variable ViGAMMAX}$<0$ (thanks to F.~Filthaut).

$\bullet$~A bug has been fixed, which affected the computation of 
branching ratios in $t\bar{t}$ and single-top production; $\alpha_{em}(q^2)$
was previously called with argument $m_{top}$ rather than $m_{top}^2$.
This only affects event weights (i.e. not distributions), and is
numerically very small.

$\bullet$~A bug in \HW\ versions 6.500 -- 6.510 can lead to occasional
violation of momentum conservation when the \HW\ parameter
{\code PRESPL=.FALSE.} (hard subprocess rapidity preserved), as is
formally assumed in MC@NLO.  Therefore at present we leave this
parameter at its default value, {\code PRESPL=.TRUE.} (hard subprocess 
longitudinal momentum preserved).  We have checked that  this formal
inconsistency has negligible actual consequences.  The bug will be
fixed in \HW\ version 6.520; meanwhile, the fix may be found on the
Fortran \HW\ wiki at
http://projects.hepforge.org/fherwig/trac/report (ticket 33).  When
this fix is implemented, the statement {\code PRESPL=.FALSE.} must be
inserted in {\code mcatnlo\_hwdriver.f} at the place indicated by the
comments therein.

$\bullet$~It has been found that a simpler form for the MC subtraction
terms with respect to that of eq.~(B.43) of ref.~\cite{Frixione:2003ei}
can be adopted; this form is now implemented in version 3.4.
This change is relevant only to $Q\bar{Q}$ and single-top
production, since for the other processes the new form and that of 
eq.~(B.43) (which is implemented in MC@NLO version 3.3 or earlier)
coincide. The differences between the two forms are equivalent 
to power-suppressed terms. This has been verified 
by comparing results obtained with version 3.4 for $t\bar{t}$ 
and single-top ($s$- and $t$-channel) production at the Tevatron and 
the LHC, and for $b\bar{b}$ production at the Tevatron, with analogous
results obtained with version 3.3. On the other hand, $b\bar{b}$ production 
at the LHC does display large differences, owing to the fact that 
the old form of MC subtraction terms has a pathology which affects
this process. Starting from version 3.4 $b\bar{b}$ production 
at the LHC may be considered safe. Technical details on the new form
of the MC subtraction terms will be posted on the MC@NLO web page.

\section{Running the package without the shell scripts\label{app:instr}}
In this appendix, we describe the actions that the user needs to 
take in order to run the package without using the shell scripts,
and the {\variable Makefile}. Examples are given for vector boson
pair production, but only trivial modifications are necessary in
order to treat other production processes.

\subsection{Creating the executables\label{app:exe}}
An $\MCatNLO$ run requires the creation of two executables, for the \NLO\
and \MC\ codes respectively. The files to link depend on whether one
uses PDFLIB, LHAPDF, or the PDF library provided with this package; 
we list them below:
\begin{itemize}
\item {\bf NLO with private PDFs:}
{\code mcatnlo\_vbmain.o mcatnlo\_vbxsec.o mcatnlo\_helas2.o 
mcatnlo\_date.o mcatnlo\_int.o mcatnlo\_uxdate.o mcatnlo\_uti.o 
mcatnlo\_str.o mcatnlo\_pdftomlm.o mcatnlo\_libofpdf.o dummies.o SYSFILE}
\item {\bf NLO with PDFLIB:}
{\code mcatnlo\_vbmain.o mcatnlo\_vbxsec.o mcatnlo\_helas2.o 
mcatnlo\_date.o mcatnlo\_int.o mcatnlo\_uxdate.o mcatnlo\_uti.o 
mcatnlo\_str.o mcatnlo\_mlmtopdf.o dummies.o}
{\variable SYSFILE CERNLIB}
\item {\bf NLO with LHAPDF:}
{\code mcatnlo\_vbmain.o mcatnlo\_vbxsec.o mcatnlo\_helas2.o 
mcatnlo\_date.o mcatnlo\_int.o mcatnlo\_uxdate.o mcatnlo\_lhauti.o 
mcatnlo\_str.o mcatnlo\_mlmtolha.o dummies.o}
{\variable SYSFILE LHAPDF}
\item {\bf MC with private PDFs:}
{\code mcatnlo\_hwdriver.o mcatnlo\_hwlhin.o mcatnlo\_hwanvbp.o 
mcatnlo\_hbook.o mcatnlo\_str.o mcatnlo\_pdftomlm.o mcatnlo\_libofpdf.o 
dummies.o} {\variable HWUTI HERWIGVER}
\item {\bf MC with PDFLIB:}
{\code mcatnlo\_hwdriver.o mcatnlo\_hwlhin.o mcatnlo\_hwanvbp.o 
mcatnlo\_hbook.o mcatnlo\_str.o mcatnlo\_mlmtopdf.o dummies.o}
{\variable HWUTI HERWIGVER CERNLIB}
\item {\bf MC with LHAPDF:}
{\code mcatnlo\_hwdriver.o mcatnlo\_hwlhin.o mcatnlo\_hwanvbp.o 
mcatnlo\_hbook.o mcatnlo\_str.o mcatnlo\_mlmtolha.o dummies.o}
{\variable HWUTI HERWIGVER LHAPDF}
\end{itemize}
The process-specific codes {\code mcatnlo\_vbmain.o} and
{\code mcatnlo\_vbxsec.o} (for the \NLO\ executable) and
{\code mcatnlo\_hwanvbp.o} (the \HW\ analysis routines in the
\MC\ executable) need to be replaced by their analogues for 
other production processes, which can be easily read from the
list given in sect.~\ref{sec:packfile}.

The variable {\variable SYSFILE} must be set either equal to {\code alpha.o},
or to {\code linux.o}, or to {\code sun.o}, according to the architecture 
of the machine on which the run is performed. For any other architecture,
the user should provide a file corresponding to {\code alpha.f} etc.,
which he/she will easily obtain by modifying {\code alpha.f}. The 
variables {\variable HWUTI} and {\variable HERWIGVER} have been described
in sect.~\ref{sec:scrvar}. In order to create the object files eventually 
linked, static compilation is always recommended (for example, 
{\code g77 -Wall -fno-automatic} on Linux).

\subsection{The input files\label{app:input}}
Here, we describe the inputs to be given to the \NLO\ and 
\MC\ executables in the case of vector boson pair production. The case
of other production processes is completely analogous.
When the shell scripts are used to run the $\MCatNLO$,
two files are created, {\variable FPREFIXNLOinput} and 
{\variable FPREFIXMCinput}, which are read by the \NLO\ and \MC\ executable
respectively. We start by considering the inputs for the \NLO\
executable, presented in table~\ref{tab:NLOi}.
\begin{table}[htb]
\begin{center}
\begin{tabular}{ll}
\hline
 '{\variable FPREFIX}'                       & ! prefix for BASES files\\
 '{\variable EVPREFIX}'                      & ! prefix for event files\\
  {\variable ECM FFACT FREN FFACTMC FRENMC}  & ! energy, scalefactors\\
  {\variable IPROC}                        & ! -2850/60/70/80=WW/ZZ/ZW+/ZW-\\
  {\variable WMASS ZMASS}                    & ! M\_W, M\_Z\\
  {\variable UMASS DMASS SMASS CMASS BMASS GMASS} & ! quark and gluon masses\\
 '{\variable PART1}'  '{\variable PART2}'    & ! hadron types\\
 '{\variable PDFGROUP}'   {\variable PDFSET} & ! PDF group and id number\\
  {\variable LAMBDAFIVE}                     & ! Lambda\_5, $<$0 for default\\
 '{\variable SCHEMEOFPDF}'                   & ! scheme\\
  {\variable NEVENTS}                        & ! number of events\\
  {\variable WGTTYPE}                 & ! 0 =$>$ wgt=+1/-1, 1 =$>$ wgt=+w/-w\\
  {\variable RNDEVSEED}                      & ! seed for rnd numbers\\
  {\variable zi}                             & ! zi\\
  {\variable nitn$_1$ nitn$_2$}              & ! itmx1,itmx2\\
\hline\\
\end{tabular}
\end{center}
\caption{\label{tab:NLOi}
Sample input file for the \NLO\ code (for vector boson pair production). 
{\variable FPREFIX} and {\variable EVPREFIX} must be understood with 
{\variable SCRTCH} in front (see sect.~\ref{sec:scrvar}).
}
\end{table}
The variables whose name is in uppercase characters have been described 
in sect.~\ref{sec:scrvar}. The other variables are assigned by the shell
script. Their default values are given in table~\ref{tab:defNLO}.
\begin{table}[htb]
\begin{center}
\begin{tabular}{ll}
\hline
Variable & Default value\\
\hline
{\variable zi}          & 0.2\\
{\variable nitn$_i$}    & 10/0 ({\variable BASES=ON/OFF})\\
\hline\\
\end{tabular}
\end{center}
\caption{\label{tab:defNLO}
Default values for script-generated variables in {\code FPREFIXNLOinput}.
}
\end{table}
Users who run the package without the script should use the values
given in table~\ref{tab:defNLO}. The variable {\variable zi} controls,
to a certain extent, the number of negative-weight events generated 
by the $\MCatNLO$ (see ref.~\cite{Frixione:2002ik}). Therefore, the user
may want to tune this parameter in order to reduce as much as possible
the number of negative-weight events. We stress that the \MC\ code will
not change this number; thus, the tuning can (and must) be done only 
by running the \NLO\ code. The variables {\variable nitn$_i$} control
the integration step (see sect.~\ref{sec:evfile}), which can be
skipped by setting {\variable nitn$_i=0$}. If one needs to perform the
integration step, we suggest setting these variables as indicated in
table~\ref{tab:defNLO}. 

\begin{table}[htb]
\begin{center}
\begin{tabular}{ll}
\hline
 '{\variable EVPREFIX.events}'               & ! event file\\
  {\variable NEVENTS}                        & ! number of events\\
  {\variable pdftype}                      & ! 0-$>$Herwig PDFs, 1 otherwise\\
 '{\variable PART1}'  '{\variable PART2}'    & ! hadron types\\
  {\variable beammom beammom}                & ! beam momenta\\
  {\variable IPROC}                         & ! --2850/60/70/80=WW/ZZ/ZW+/ZW-\\
 '{\variable PDFGROUP}'                      & ! PDF group (1)\\
  {\variable PDFSET}                         & ! PDF id number (1)\\
 '{\variable PDFGROUP}'                      & ! PDF group (2)\\
  {\variable PDFSET}                         & ! PDF id number (2)\\
  {\variable LAMBDAHERW}                     & ! Lambda\_5, $<$0 for default\\
  {\variable WMASS WMASS ZMASS}              & ! M\_W+, M\_W-, M\_Z\\
  {\variable UMASS DMASS SMASS CMASS BMASS GMASS} & ! quark and gluon masses\\
\hline\\
\end{tabular}
\end{center}
\caption{\label{tab:MCi}
Sample input file for the \MC\ code (for vector boson pair production), 
resulting from setting {\variable HERPDF=EXTPDF}, which implies 
{\variable pdftype=1}. 
Setting {\variable HERPDF=DEFAULT} results in an analogous file, with
{\variable pdftype=0}, and without the lines concerning
{\variable PDFGROUP} and {\variable PDFSET}. {\variable EVPREFIX} 
must be understood with {\variable SCRTCH} in front 
(see sect.~\ref{sec:scrvar}). The negative sign of {\variable IPROC}
tells \HW\ to use Les Houches interface routines.
}
\end{table}
We now turn to the inputs for the \MC\ executable, presented
in table~\ref{tab:MCi}. 
The variables whose names are in uppercase characters have been described 
in sect.~\ref{sec:scrvar}. The other variables are assigned by the shell
script. Their default values are given in table~\ref{tab:defMC}.
\begin{table}[htb]
\begin{center}
\begin{tabular}{ll}
\hline
Variable & Default value\\
\hline
{\variable esctype}         & 0\\
{\variable pdftype}         & 0/1 ({\variable HERPDF=DEFAULT/EXTPDF})\\
{\variable beammom}         & {\variable EMC}/2\\
\hline\\
\end{tabular}
\end{center}
\caption{\label{tab:defMC}
Default values for script-generated variables in {\code MCinput}.
}
\end{table}
The user can freely change the values of {\variable esctype} and
{\variable pdftype}; on the other hand, the value of {\variable beammom}
must always be equal to half of the hadronic CM energy.

When LHAPDF is linked, the value of {\variable PDFSET} is sufficient
to identify the parton density set. In such a case, {\variable PDFGROUP}
must be set in input equal to {\variable LHAPDF} if the user wants
to freeze the PDFs at the boundaries (defined as the ranges in which
the fits have been performed). If one chooses to extrapolate the PDFs
across the boundaries, one should set {\variable PDFGROUP=LHAEXT}
in input.

In the case of $\gamma/Z$, $W^\pm$, Higgs or heavy quark production, the 
\MC\ executable can be run with the corresponding positive input process 
codes {\variable IPROC} = 1350, 1399, 1499, 1600+ID, 1705, 1706,
2000--2008, 2600+ID or 2700+ID,
to generate a standard \HW\ run for comparison purposes\footnote{For
vector boson pair production, for historical reasons, the different
process codes 2800--2825 must be used.}.  Then the input
event file will not be read: instead, parton configurations will be
generated by \HW\ according to the LO matrix elements.


\end{document}